# Selective observation of Goos–Hänchen and Imbert–Fedorov shifts in partial reflection via optimized weak measurements in linear and elliptical polarization basis

**S. Goswami[1,2], S. Dhara[1], M. Pal[1], A. Nandi[1,3], P. K. Panigrahi[1] and N. Ghosh[1,*]**

[1]Department of Physical Sciences, Indian Institute of Science Education and Research-Kolkata, Mohanpur Campus, Mohanpur 741 252, India
[2]Department of Physics & Astronomy, University Of Calgary, 2500 University Drive NW, Calgary, AB T2N 1N4, Canada
[3] Department of Physics and Astronomy, Purdue University,,West Lafayette, IN 47907-2036, USA
[*] nghosh@iiserkol.ac.in

**Abstract:** The spatial and the angular variants of the Goos-Hänchen (GH) and the Imbert-Federov (IF) beam shifts contribute in a complex interrelated way to the resultant beam shift in partial reflection at planar dielectric interfaces. Here, we show that the angular GH and the two variants of the IF effects can be decoupled, amplified and separately observed by weak value amplification and subsequent conversion of spatial ↔ angular nature of the beam shifts using appropriate pre and post selection of polarization states. We experimentally demonstrate such decoupling and illustrate various other intriguing manifestations of weak measurements by employing optimized pre and post selections (based on the eigen polarization states of the shifts) elliptical and / or linear polarization basis. The demonstrated ability to amplify, controllably decouple or combine the beam shifts via weak measurements may prove to be valuable for understanding the different physical contributions of the effects and for their applications in sensing and precision metrology.

---

## References and links

1. F. Goos and H. Hänchen, "Ein neuer und fundamentaler Versuch zur Totalreflexion," Annals of Physics, 1, 333 (1947).
2. C. Imbert, "'Calculation and Experimental Proof of the Transverse Shift Induced by Total Internal Reflection of a Circularly Polarized Light Beam'',Physical Review D, **5**, 787 (1972).
3. A. Aiello, "Goos–Hänchen and Imbert–Fedorov shifts: a novel perspective" New Journal of Physics., **14**, 013058 (2012).
4. *K.Y. Bliokh, A Aiello, "Goos–Hänchen and Imbert–Fedorov beam shifts: an overview'' Journal of Optics, 15, 014001 (2013).*
5. O. Hosten and P. Kwiat, "Observation of the Spin Hall Effect of Light via Weak Measurements",Science, **319**, 787 (2008)
6. M. Merano, A. Aiello, M.P. van Exter and J. P. Woerdman, "Observing angular deviations in the specular reflection of a light beam" Nature Photonics., **3**, 337 (2009).
7. A. Aiello and J. P. Woerdman, "Role of beam propagation in Goos–Hänchen and Imbert–Fedorov shifts" Optics Letters., **33**, 1437 (2008).
8. Y Aharonov, DZ Albert, L Vaidman, "How the result of a measurement of a component of the spin of a spin-1/2 particle can turn out to be 100", Physical Review Letter, 60, 1351 (1988).
9. F. Toppel, M. Ornigotti, and A. Aiello, "Goos–Hanchen and Imbert–Fedorov shifts from a quantum-mechanical perspective" New Journal of Physics., **15**, 113059 (2013).

10. J.B. Gotte, M.R. Dennis, "Generalized shifts and weak values for polarization components of reflected light beams", New Journal of Physics., **14**, 073016 (2012).

11. S Goswami, M Pal, A Nandi, PK Panigrahi and N Ghosh, "Simultaneous weak value amplification of angular Goos–Hänchen and Imbert–Fedorov shifts in partial reflection", Optics Letter, 39, 6229 (2014).

12. G Jayaswal, G Mistura, and M Merano, "Observing angular deviations in light-beam reflection via weak measurements", Optics Letter, **39**, 6257 (2014)

13. G. Jayaswal, G. Mistura and M. Merano, "Weak measurement of the Goos–Hänchen shift", Optics Letters., **38**, 1232 (2013).

14. G. Jayaswal, G. Mistura and M. Merano, "Observation of the Imbert–Fedorov effect via weak value amplification" Optics Letters., **39**, 2266 (2014).

15. A. Aiello A and J. P. Woerdman, "Theory of angular Goos-Hänchen shift near Brewster incidence", arXiv: 0903.3730 (2009).

## 1.Introduction

Laws of Geometrical optics govern the partial reflection/refraction or total internal reflection (TIR) of plane waves. However, finite or bounded light beams do not follow the same [1-4]. Finite beams such as the fundamental or the higher order Gaussian beams exhibit shifts in their centroid in both longitudinal (in the plane of incidence) or transverse (perpendicular to the plane of incidence) directions. The longitudinal shift is called Goos–Hänchen (GH) shift and the transverse one is known as Imbert–Fedorov (IF) shift [1-2]. Although, these effects can be interpreted to occur due to the interference of the constituent plane waves (with modified amplitudes and / or phase as a result of the interaction) in a finite beam, they have profoundly different physical origins. While, the GH shift owes its origin to the angular gradient of the complex reflection / refraction coefficients ($r_p$ and $r_s$), the IF shift (also called Spin Hall(SH) shift of light [4-5]) originates from the evolution of geometric phase and the resulting spin orbit interaction of light [1-4]. The eigen polarization modes of the GH shift are TM (p) and TE (s) linear polarizations and that of the IF shift are both left/right circular (elliptical) polarizations and diagonal ($\pm 45^\circ$) linear polarizations. Note that in the case of TIR, both the GH shift and the two variants of the IF shifts are spatial in nature (coordinate shift manifested as spatial displacement) [4]. In case of partial (non-total) reflection, on the other hand, the different variants of the GH and the IF shifts contribute in a more complex interrelated way to the resultant beam shift. Here, the GH shift is angular in nature (momentum domain shift manifested as angular deflection) and the IF shift can either be spatial (for circular/elliptical polarizations) or angular (for $\pm 45^\circ$ linear polarizations) [4].

These beam shifts are under recent intensive investigations because of their fundamental nature and potential applications [4-7]. However, the exceedingly small magnitude of the shifts (typically in the sub-wavelength domain) is a major stumbling block towards their practical applications. Weak measurement, despite being discovered in the context of quantum mechanics [8], is applicable in the context of the beam shifts in classical optics [7,9-10], and have thus been employed recently to amplify and reliably observe the tiny optical beam shifts using conventional diffraction limited detection system [5-6,9,11-14]. However, as previously noted, the different variants of the GH and the IF shifts contribute in a complex interrelated way in partial reflection. Decoupling and selective weak value amplification of the contributing effects (GH and the two variants of the IF) with appropriate weak measurement schemes should prove to be valuable for understanding the different physical contributions of the shifts and for their unique interpretation. Moreover, this may also be helpful towards designing/optimizing schemes for their applications in sensing and precision metrology. In this paper, we have therefore addressed this issue by selectively observing the shifts through the conversion of spatial ↔angular nature of the beam shifts (or by retaining its original angular nature) via weak measurements. This is accomplished by suitable choosing pre and post selection of (elliptical and / or linear) polarization states, based on the eigen polarization modes of the different shifts. An angular shift can be manifold amplified compared to a spatial shift by increasing the beam propagation distance and reducing the beam waist (resulting shift is inversely proportional to the square of the beam waist) [4,7]. Here, we describe two such different sets of modified weak measurement schemes for the selective amplification of the GH and the IF shifts in partial reflection. In the first set, we employ pre-selection in circular polarization (LCP/RCP) basis and post-selection in two different elliptical polarization basis to separately observe the weak value amplified angular GH and IF (eigen modes - $\pm 45°$ linear polarizations) shifts. In the second set of schemes, pre-selection at linear polarization (either p or s-states) and post-selections at linear and elliptical polarization basis are used for weak value amplification and selective observation of the IF shifts. We provide interesting examples of weak measurements by employing post-selections in two different elliptical polarization basis, wherein in one case, all the shifts are converted to spatial nature leading to no observable beam shifts. In the other case, the two variants of the IF effects combine in an intriguing fashion to yield partially spatial and partially angular weak value amplified IF shifts. These intriguing manifestations of beam shifts predicted by the theoretical treatment of weak measurements corroborate well with experimental results.

## Theory

The GH and IF shifts of fundamental Gaussian beam can be described by polarization operators [4]

$$GH = \begin{bmatrix} \Omega_p(\theta) & 0 \\ 0 & \Omega_s(\theta) \end{bmatrix} \ IF = \begin{bmatrix} 0 & \Omega_l(\theta) \\ -\Omega_r(\theta) & 0 \end{bmatrix} \tag{1}$$

where,

$$\Omega_p(\theta) = -i\frac{\partial \ln r_p}{\partial \theta}, \ \Omega_s(\theta) = -i\frac{\partial \ln r_s}{\partial \theta}$$

$$\Omega_l(\theta) = i\left(1 + \frac{r_p}{r_s}\right)\cot\theta, \ \Omega_r(\theta) = i\left(1 + \frac{r_s}{r_p}\right)\cot\theta \tag{2}$$

Here, $r_p(\theta)$ and $r_s(\theta)$ are Fresnel reflection coefficients for p and s linear polarizations (real in case of partial reflection), respectively and θ denotes the angle of incidence. For pre $\left|\psi_{pre}\right\rangle$ and post-selection $\left|\psi_{post}\right\rangle$ of states, the weak value of the shifts are given by

$$A_w^{GH} = \frac{\left\langle \psi_{post} \left| GH \right| \psi_{pre} \right\rangle}{\left\langle \psi_{post} \middle| \psi_{pre} \right\rangle} \text{ and } A_w^{IF} = \frac{\left\langle \psi_{post} \left| IF \right| \psi_{pre} \right\rangle}{\left\langle \psi_{post} \middle| \psi_{pre} \right\rangle} \tag{3}$$

Note that the pre-selected state is not the input polarization state $\left|\psi_{in}\right\rangle$ here, rather it is $\left|\psi_{pre}\right\rangle = R\left|\psi_{in}\right\rangle$. Here, R is a 2×2 diagonal matrix (with its elements $r_p(\theta)$ and $r_s(\theta)$) representing the Fresnel reflection Jones matrix. We now proceed to our proposed modified weak measurement schemes.

**Scheme-1:** *Pre-selection in circular polarization (LCP/RCP) and post-selection in nearly orthogonal elliptical polarization states*

The input polarization state is chosen $\left|\psi_{in}\right\rangle \approx [\alpha_1, \alpha_2]^T \approx \left[\frac{1}{r_p}, i\frac{1}{r_s}\right]^T$ such that the pre-selected polarization state becomes LCP ($\left|\psi_{pre}\right\rangle \approx [1, i]^T$ ).Here and henceforth, we have used two component (un-normalized) Jones vector to represent the polarization states [4]. In weak

measurements, the post selected state $\left|\psi_{post}\right\rangle$ is nearly (but not exactly) orthogonal to $\left|\psi_{pre}\right\rangle$, leading to a large magnitude of the weak value ($A_w$ of Eq. 3). In this case, the exactly orthogonal RCP state $\left[1,-i\right]^{\mathrm{T}}$ can be produced in infinite different ways using a combination of a quarter waveplate (QWP) and a linear polarizer in sequence, by orienting the axis of QWP at an angle $45^{\mathrm{O}}$ with respect to the polarizer axis. The post-selections at nearly orthogonal elliptical polarization states can then be done by rotating the polarizer a small angle $\pm\varepsilon$ off from the exact orthogonal position. In the following, we discuss two such different post-selection schemes, to separately observe the angular GH and IF (eigen modes $\pm45^\circ$ linear polarizations) shifts. Note that in this case, there would be no weak measurements on the spatial IF shift with LCP/RCP eigen polarization states, as the pre-selected state is LCP.

(a): *QWP axis kept along the vertical direction, followed by a polarizer with its axis at an angle $-(45^{\mathrm{O}}\pm\varepsilon)$ with respect to the horizontal axis (direction of p-polarization):*

$$\left|\psi_{post}\right\rangle \approx \begin{bmatrix} 1\mp\varepsilon \\ -i(1\pm\varepsilon) \end{bmatrix}$$

The expressions for the weak values of GH and IF shifts can be derived using Eq. (3) as

$$A_w^{a,GH} = \pm i\frac{\left(\dfrac{r_p'}{r_p}-\dfrac{r_s'}{r_s}\right)}{2\varepsilon} - \frac{i}{2}\left(\frac{r_p'}{r_p}+\frac{r_s'}{r_s}\right), \quad A_w^{a,IF} = \pm\frac{\left(\dfrac{r_p}{r_s}-\dfrac{r_s}{r_p}\right)}{2\varepsilon}\cot\theta - \frac{1}{2}\left(\frac{r_s}{r_p}+\frac{r_p}{r_s}+2\right)\cot\theta \quad (4)$$

Henceforth, the $\pm$ signs correspond to post selections with $\pm\varepsilon$ and the superscripts (*a, b, c* etc.) correspond to post-selection schemes.

(b): *QWP axis kept at an angle $45^{\mathrm{O}}$ with respect to the horizontal (direction of p-polarization), followed by a polarizer at $90^\circ \mp \varepsilon$ with respect to the horizontal axis:*

$$\left|\psi_{post}\right\rangle \approx \begin{bmatrix} (1+i)\pm\varepsilon(1-i) \\ (1-i)\pm\varepsilon(1+i) \end{bmatrix}$$

The corresponding weak values are

$$A_w^{b,GH} = \mp\frac{1}{2\varepsilon}\left(\frac{r_p'}{r_p}-\frac{r_s'}{r_s}\right) - \frac{i}{2}\left(\frac{r_p'}{r_p}+\frac{r_s'}{r_s}\right), \quad A_w^{a,IF} = \pm\frac{i}{2\varepsilon}\left(\frac{r_p}{r_s}-\frac{r_s}{r_p}\right)\cot\theta - \frac{1}{2}\left(\frac{r_s}{r_p}+\frac{r_p}{r_s}+2\right)\cot\theta \quad (5)$$

Note, real and imaginary weak values correspond to shifts in co-ordinate (spatial shift) and momentum (angular shift) spaces, respectively [4-5,8-10]. The angular shift (having imaginary weak value) gets coupled to the beam propagation and with appropriate choice of the beam parameters (beam waist and Rayleigh range) and propagation distance. This can be made several orders of magnitude larger than the corresponding spatial shift (which is independent of the beam waist and the propagation distance). As apparent from Eq. (4), with post-selection scheme (a), the GH shift weak value is purely imaginary and hence the weak value amplified shift (the part $\propto \dfrac{1}{\varepsilon}$) is angular, whereas the IF shift weak value is real and hence the amplified shift is spatial in nature. Thus, with this post-selection, one can decouple the IF effect and experimentally observe exclusively the GH shift. In case of post-selection with scheme (b), on the other hand, one can experimentally observe the weak value amplified IF shift having $\pm45^\circ$ linear polarizations as eigen modes (imaginary part of Eq. 5, which yields the weak value amplified angular shift) by decoupling the GH shift. The imaginary parts of the weak value amplified shifts in Eqs. (4) and (5) would be manifested as amplified angular shifts $\left\langle\Delta\theta\right\rangle_w^{a,GH}$ and $\left\langle\Delta\theta\right\rangle_w^{b,IF}$ between the two post selected elliptical polarization states ($\pm\varepsilon$ away from the orthogonal RCP state) either in scheme (a) or (b) respectively. This eventually manifests as observable shifts in the centroid of a Gaussian beam in a direction parallel ($\left\langle\Delta x\right\rangle^{GH}$, case-(a)) and perpendicular ($\left\langle\Delta y\right\rangle^{IF}$, case-(b)) to the plane of incidence, respectively. The corresponding expressions can be obtained as [11]

$$\left\langle\Delta\theta\right\rangle_w^{a,GH}=\frac{\lambda}{\pi\varepsilon z_0}\left(\frac{r_p'}{r_p}-\frac{r_s'}{r_s}\right),\ \left\langle\Delta x\right\rangle_a^{GH}=z\left\langle\Delta\theta\right\rangle_w^{a,GH} \tag{6a}$$

$$\left\langle\Delta\theta\right\rangle_w^{b,IF}=\frac{\lambda}{\pi\varepsilon z_0}\left(\frac{r_p}{r_s}-\frac{r_s}{r_p}\right)\cot\theta,\ \left\langle\Delta y\right\rangle_b^{IF}=z\left\langle\Delta\theta\right\rangle_w^{b,IF} \tag{6b}$$

Here, $r_p'\ (r_s')$ are angular derivatives of $r_p\ (r_s)$, $\lambda$ is the wavelength, $z$ is the propagation distance and $z_0$ is the Rayleigh range of a fundamental Gaussian beam. As evident, by varying the post-selected elliptical polarization state ($\pm\varepsilon$ away from the RCP) only slightly one could turn an angular shift into a spatial one or vice-versa (spatial $\leftrightarrow$ angular), or could retain its original nature.

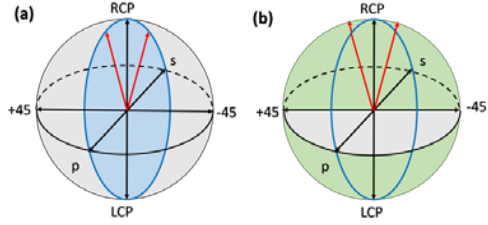

Fig. 1. Pictorial representation of the post-selection schemes (a) and (b) in Poincare sphere. In both cases, the pre-selected state is LCP (South Pole). The eigen-states of GH (p and s linear polarization) and angular IF (+45⁰/-45⁰ linear polarizations) shift (black arrows) and the post-selected elliptical polarization states (red arrows) are shown. Two separate planes, one containing the pre-selected state and the eigen-states of GH (blue shadow), the other containing the pre-selected state and the eigen-states of angular IF (green shadow) are marked in both the figures.

Some useful insights on the effect of post-selection can be gained by representing the states in the Poincaré sphere (wherein the polarization states are represented by Stokes vector rather than the Jones vector). As shown in Fig. 1, for post-selection scheme (a), the pre-selected state (LCP), the post-selected states ($\pm\varepsilon$ away from the RCP) and the eigen-states of angular GH shift (p, s linear polarizations) are in the same plane (Fig. 1a). In contrast, the post-selected states are out of the plane formed by the eigen-states of angular IF shift ($\pm45°$ linear polarizations) and the pre-selected state (LCP). This post selection results in the retaining of the original angular nature of the weak value amplified GH shift, whereas the angular IF shift is converted to spatial shift in the weak measurements. The situation is exactly reversed in scheme-(b), where the pre-selected state and the post-selected states are in the same plane with angular IF shift eigen-states but not with the angular GH shift eigen-states (Fig. 1b). Accordingly, in this case, the weak value amplified IF shift (having $\pm45°$ linear polarizations as eigen-modes) retains its original angular nature and the angular GH shift is converted to spatial in the weak measurements. A general rule governing this can be stated in a simple form. If the pre-selected state, the post-selected states and the eigen-states of a particular shift are in the same plane then after weak value amplification, the original nature of the shift

(spatial or angular) is retained, whereas if the post-selected states are in a perpendicular plane to the others then the nature of weak value amplified shift is converted (angular to spatial or vice versa). Continuing on the role of the post-selection, we add that if the post-selected states are in intermediate angles, neither in-plane or perpendicular to the plane formed by the pre-selected state and the eigen-states of a given shift, the weak value amplified shift may become partially spatial and partially angular, as we address next.

**Scheme-2:** *Pre-selection in p/s linear polarization and post-selection in nearly orthogonal linear / elliptical polarization*

Here, the pre-selection states are either p $\left|\psi_{pre}\right\rangle \approx [1,0]^T$. or s $\left|\psi_{pre}\right\rangle \approx [0,1]^T$ linear polarizations, leading to no weak value amplifications of the angular GH effect. We consider three different post-selection schemes – (a) in linear polarization basis, (b) and (c) in elliptical polarization basis. Post-selection using scheme (a) has been addressed previously [11], and hence not discussed in details here. The weak value difference between ±ε states for the IF shift in this case is known to be given by [11]

$$\Delta A_{w, p/s}^{a, IF} = \frac{2i}{\varepsilon}\left(1 + \frac{r_{s/p}}{r_{p/s}}\right)\cot\theta \qquad (7)$$

Here onwards, the subscripts 'p/s' correspond to pre-selected states at either p or s polarizations. Here, in order to avoid confusion regarding the ± signs in the actual weak values (arising due to ±ε states), we have resorted to the differential weak value $\Delta A_w (= A_w|_{+\varepsilon} - A_w|_{-\varepsilon})$, which is pertinent to the experimental shift. We now discuss scheme (b) and (c) dealing with post-selection in nearly orthogonal (to $\left|\psi_{pre}\right\rangle$) elliptical polarization states. Once again, using a combination of a QWP and a linear polarizer in sequence, two different post-selection schemes can be implemented.

*(b): QWP axis oriented at an angle ±ε with respect to the axis of a polarizer (placed after QWP) which is oriented at 90$^O$ with respect to the pre-selected linear polarization (either p or s):*

Here, $\left|\psi_{post}\right\rangle \approx \begin{bmatrix} \mp\varepsilon(1+i) \\ 1 \end{bmatrix}$ and $\begin{bmatrix} 1 \\ \pm\varepsilon(1-i) \end{bmatrix}$ for p and s pre-selections, respectively.

The expressions for the weak values of IF shift can be obtained as

$$\Delta A_{w, p/s}^{b, IF} = \frac{(1\pm i)i}{\varepsilon}\left(1 + \frac{r_{s/p}}{r_{p/s}}\right)\cot\theta \qquad (8)$$

*(c): QWP axis oriented at an angle 90$^O$ with respect to the pre-selected linear polarization (either p or s) followed by a polarizer oriented at 90$^O$±ε with respect to the pre-selected linear polarizations:*

Here, $\left|\psi_{post}\right\rangle \approx \begin{bmatrix} \mp i\varepsilon \\ 1 \end{bmatrix}$ and $\begin{bmatrix} 1 \\ \mp i\varepsilon \end{bmatrix}$ for p and s pre-selections, respectively.

The expressions for thecorresponding weak values of IF shift are

$$\Delta A_{w, p/s}^{c, IF} = \mp\frac{2}{\varepsilon}\left(1 + \frac{r_{s/p}}{r_{p/s}}\right)\cot\theta \qquad (9)$$

It is evident from Eq. (8) that for post-selection scheme (b), the weak value turns out to be complex with equal magnitudes of real and imaginary parts, implying partial spatial and partial angular weak value amplified shifts. Here, the post-selected states are neither in-plane nor perpendicular to the planes formed by the pre-selected state and the eigen-states of the two variants of the IF effect (spatial IF with LCP/RCP eigen-modes and angular IF with ±45° polarizations eigen-modes). Hence, as previously noted, the resulting weak value amplified angular IF shift is contributed by both the variants of the IF effects with partial conversion and partial retaining of their original nature. For scheme (c), on the other hand, the weak value is purely real leading to completely spatial weak value amplified IF shift. This can also be explained by the position of the post-selected states in the Poincaré sphere exactly analogous to Figure 1.

The experimentally observable shifts in the beam centroid ($\langle \Delta y \rangle_{w,p/s}^{IF}$) between two post selected polarization states ($\pm \varepsilon$ away from the orthogonal state) for the above three cases can be summarized from the corresponding weak value amplified angular part of the shifts ($\langle \Delta \theta \rangle_{w,p/s}^{a/b/c,IF}$) as

$$\langle \Delta \theta \rangle_{w,p/s}^{b,IF} = \frac{1}{2} \langle \Delta \theta \rangle_{w,p/s}^{a,IF} \ , \ \langle \Delta \theta \rangle_{w,p/s}^{c,IF} = 0 \ , \ \langle \Delta y \rangle_{w,p/s}^{IF} = z \langle \Delta \theta \rangle_{w,p/s}^{IF} \tag{10}$$

As apparent from Eq. (10), the completely angular nature of the weak value amplified IF shift in post-selection scheme (a) is turned to half angular half spatial (observable shift in the beam centroid reduced in magnitude by a factor of half) in scheme (b) to completely spatial (observable shift vanishes) in scheme (c). We now turn to experimental realization of the aforementioned weak measurement schemes (Scheme-1 and 2).

## Experiments

In our experimental system (shown in Figure 2), the 632.8 nm line of a He-Ne laser (HRR120-1, Thorlabs, USA) was used as the source. The beam was spatially filtered, collimated and then focused by a lens L (focal length f = 15 cm) to a spot size of $\omega_0 \sim 100$ μm. The polarization state of the input beam is controlled by a combination of a Glan-Thompson linear polarizer P1 (GTH10M-A, Thorlabs, USA) and a removable quarter waveplate QWP1 (WPQ10M-633, Thorlabs, USA) mounted on computer controlled precision rotational mounts (PRM1/MZ8, Thorlabs, USA). The beam then undergoes external reflection from the surface of a 45°-90°-45° BK7 prism (PS912, Thorlabs, USA, refractive index n = 1.516) mounted on a computer controlled Nano Rotation Stage (NR360S/M, Thorlabs, USA). The reflected beam is passed through the polarization post-selecting unit, comprising of a similar arrangement of rotatable Glan-Thompson linear polarizer P2 and rotatable quarter waveplate QWP2, but placed in reverse order. The resulting beam shift is detected by a CCD camera (MP3.3-RTV-R-CLR-10-C, Singapore, $2048 \times 1536$ square pixels, pixel dimension $3.45 \mu m$, $3 \times 3$ binning). The propagation distance z was chosen to be 38 cm.

In order to realize *Scheme 1* of weak measurements, the polarization post selecting unit (P2 and QWP2) was first oriented to RCP analyzer state. The two different RCP analyzer states were first obtained using two different set of orientations of the QWP2 and P2 (as per the post-selection schemes 1a and 1b) The pre-selecting P1 and QWP1 were then oriented to observe the intensity minimum. At this position, the input state is $\left( \left| \psi_{in} \right\rangle \approx \left[ \frac{1}{\sqrt{p}}, \frac{i}{\sqrt{s}} \right]^T \right)$ and the pre-selected state is LCP $\left( \left| \psi_{pre} \right\rangle \approx \left[ 1, i \right]^T \right)$. Weak measurements were then performed by changing the polarization axis of P2 to $\pm \varepsilon$ away from this (exact orthogonal pre and post selection states) position in either of the schemes (1a - corresponding Eq. 4, and 1b- corresponding Eq. 5). In order to realize *Scheme 2* of weak measurements, the quarter waveplate QWP1 was removed and pre-selections were done in either s or p-linear polarization states by orienting the polarizer P1. The three different post-selection schemes – (a) linear polarizations, (b) and (c) – in elliptical polarizations were realized by orienting QWP2 and P2 in the post-selecting unit, as previously specified. In case of scheme 2a, measurements were performed by removing QWP2 and by changing the orientation of P2 to $\pm\varepsilon$ away from the exact orthogonal position (with respect to the pre-selected p/s linear polarization states). Similarly, in scheme 2b and 2c, measurements were performed by inserting the QWP2 and by changing the orientation of QWP2 and P2 to $\pm\varepsilon$ away from the exact orthogonal positions, respectively. Once again, the exact orthogonal positions were obtained by observing the intensity minima. In all the cases, the shift in the centroid of the reflected beam between the two post selected states ($\pm\varepsilon$ away from the orthogonal) was recorded. The measurements were performed as a function of angle of incidence ($\theta = 30^o$ - $70^o$) and for varying $\varepsilon$ ( +0.5 rad to 0.7 rad). The desired control on $\varepsilon$ and $\theta$ were achieved using the computer controlled precision rotation mounts and the nano rotation stage, respectively.

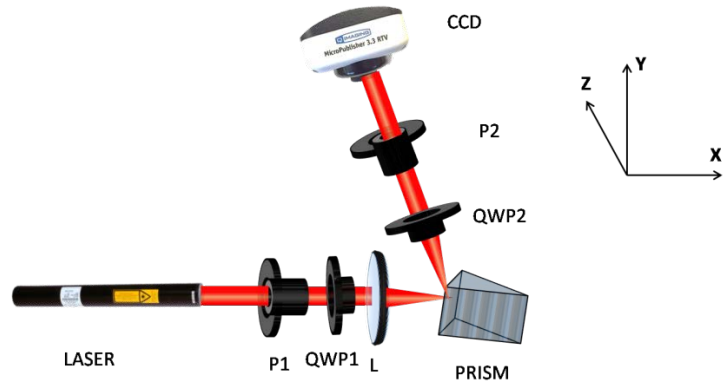

Fig. 2. A generalized schematic of the experimental system for the weak measurement of the angular GH and IF shifts in partial reflection. P1, P2: rotatable Glan-Thompson linear polarizers mounted on precision rotation mount; QWP1, QWP2 removable quarter waveplates L: Lens. The prism mounted on a precision rotation stage act as the weak measuring device.

## Results and discussion

Figure 3 illustrates the shifts in the centroid of the reflected Gaussian beam observed by employing weak measurement Scheme 1 (a and b). The results are shown for $\theta = 62^o$ and $\varepsilon = 0.7$ rad. While post selection in elliptical polarization with scheme 1(a) results in centroid shift entirely along the longitudinal (x) direction (Fig. 3a), shift in the beam centroid along the transverse (y) direction is observed by employing post-selection scheme 1b (Fig. 3b). This provides conclusive evidence of decoupling of the angular GH and the IF shifts in weak measurements employing pre and post-selection in circular (elliptical) polarization basis (scheme 1).

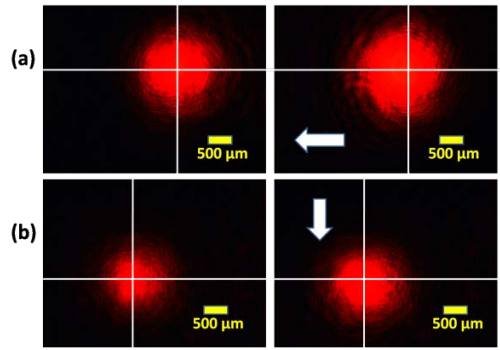

Fig. 3. Decoupling and separate weak value amplificationof angular GH and angular IF shifts in partial reflection, employing pre and post-selection in circular (elliptical) polarization basis (Scheme 1). Two different elliptical post-selection schemes (scheme 1a and 1b) are shown in (a) and (b) respectively. Longitudinal GH (along x-direction, in (a)) and transverse IF (along y-direction, in (b)) shifts in beam's centroid between the two post selected states +$\varepsilon$ (left panel) and -$\varepsilon$ (right panel) away from the orthogonal state are apparent.

The resulting dependence on the weak value amplified angular GH shift (in scheme 1a) and the angular IF shift (in scheme 1b) on the angle of incidence ($\theta$) are shown in Figure 4a and 4b respectively. Here, the physical shifts of the beam centroid ($\left\langle \Delta x \right\rangle^{GH}$, and $\left\langle \Delta y \right\rangle^{IF}$) are shown, rather than the actual angular shifts ($_{\left\langle \Delta\theta \right\rangle}^{GH}$, and $_{\left\langle \Delta\theta \right\rangle}^{IF}$). The experimental shifts are observed to be in excellent agreement with the corresponding theoretical predictions (Eqs. 6a and 6b). The observed reversal of sign (shift in opposite direction) of both the GH and the IF shifts across the Brewster angle ($\theta_B$) is consistent with the phase change of $\pi$ of the reflection coefficient $r_p$ across $\theta_B$ ($\dfrac{r_s}{r_p}$ is negative (positive) for $\theta < \theta_B$ ($\theta > \theta_B$)) and that in this angular range $|r_p| << |r_s|$. Even though the theoretical expressions of Eq. 6a and 6b are approximate (obtained by retaining the first order terms in the Taylor series expansion of Fresnel reflection coefficients), they predict the shifts with reasonable accuracy when one is sufficiently away from the Brewster angle (which was ensured in our experiments) [15]. Importantly, the excellent agreement between the theoretical predictions and the experimental shifts underscore – *(i)* successful decoupling and faithful weak value amplification of the angular GH and IF shifts, and (ii) unlike previously employed weak measurement schemes, pre and post-selection in circular (elliptical) polarization basis (in our modified weak measurement scheme 1) enables exclusive observation of the angular IF effect having $\pm 45^\circ$ linear polarizations as eigen-modes (thus decoupling it from the other variant, spatial IF effect with LCP/RCP eigen-modes).

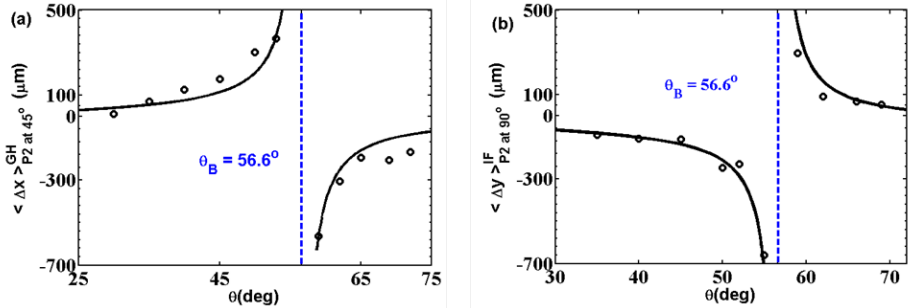

Fig. 4. The dependence of the shift of the beam centroid representing **(a)** the angular GH shift $\left\langle \Delta x \right\rangle^{GH}$ and **(b)** the angular IF shift $\left\langle \Delta y \right\rangle^{IF}$ on the angle of incidence $\theta$ for weak measurement schemes 1a and 1b, respectively. In both the figures, symbols (open circle) represent experimental data and the corresponding theoretical predictions (Eq. 6a and 6b, for $\varepsilon = 0.7$ rad) are shown by black lines. The agreement between the theory and experiment *is* seen to be excellent on either side of the Brewster angle ($\theta_B \sim 56.6^\circ$).

The results of weak measurements employing Scheme 2 (a, b and c) are summarized in Figure 5. As previously noted, for all the post-selecting schemes here, the weak value amplification is observed exclusively for the two variants of the IF shifts (no weak value amplifications of the angular GH effect). The angular dependence of the resulting weak value amplified IF shifts with three different post-selections (a: linear polarization, b and c: elliptical polarizations) for pre-selection with p-linear polarization state (Fig. 5a) and s-linear polarization state (Fig. 5a) are shown in the figure. The corresponding theoretical predictions (Eqs. 7, 8 for post-selection scheme-a, b with $\varepsilon = 0.7$ rad and Eq. 9 for post-selection scheme-c with $\varepsilon = 0.17$ rad) are also displayed.

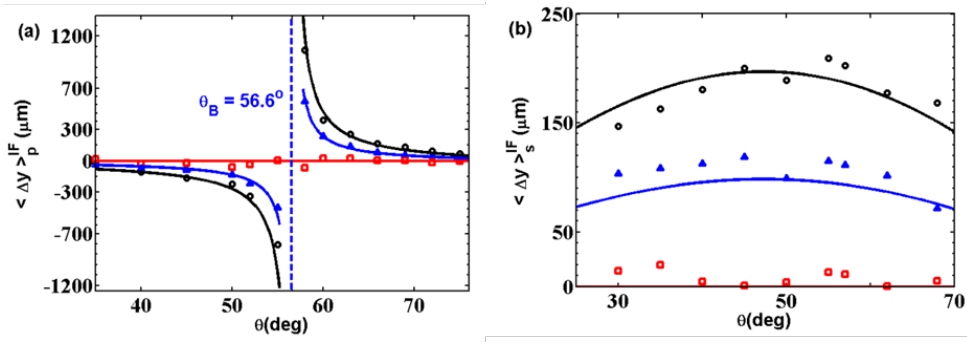

Fig. 5. The angular dependence of the weak value amplified shift of the beam centroid representing the angular IF shift $\langle \Delta y \rangle^{IF}$ for weak measurement schemes 2a, 2b and 2c. The results are shown for pre-selection with (a) p-linear polarization and (b) s-linear polarization. In both cases, the experimental IF shifts for post selections in linear polarization (scheme 2a: black circle) and two different elliptical polarizations (scheme 2b: blue triangle, scheme 2c: red square) are shown. The corresponding theoretical predictions (Eqs. 7, 8 for $\varepsilon = 0.7$ rad, and Eq. 9 for $\varepsilon = 0.17$ rad respectively) are shown by black, blue and red lines respectively.

Once again, the experimental beam shifts are in good agreement with the theoretical predictions, implying faithful weak value amplifications with the proposed schemes. These results therefore demonstrate that fully angular nature of the weak value amplified IF shift in post-selection scheme (a) is turned to half angular and half spatial (leading to a reduction in the magnitude of the observable shift in the beam centroid by a factor of half) in scheme (b) to completely spatial (the observable shift nearly vanishes) in scheme (c). As previously noted, in scheme (a) and (b), both the variants of the IF shifts (spatial IF with LCP/RCP eigen-modes and angular IF with $\pm 45°$ polarizations eigen-modes) contribute to the observed weak value amplified transverse shift by becoming angular. In scheme (a), the spatial IF is converted to angular and the angular IF retains its original angular nature. In scheme (b), on the other hand, the two variants combine in an intricate fashion with partial conversion and partial retaining of their original nature.

To summarize, we have demonstrated that the angular GH and the two variants of the IF effects having different physical origins, can be decoupled, amplified and separately observed by weak value amplification and subsequent conversion of spatial ↔ angular nature of the beam shifts using appropriate pre and post selection in elliptical and / or linear polarization basis. In our first weak measurement scheme, we have employed pre-selection in circular polarization (LCP) basis and post-selection in two different elliptical polarization basis to separately observe the weak value amplified angular GH and IF (eigen modes- $\pm 45°$ linear polarizations) shifts. Unlike previously employed weak measurement schemes, this enabled exclusive observation of the angular IF effect having $\pm 45°$ linear polarizations as eigen-modes (thus decoupling it from the other variant, spatial IF effect with LCP/RCP eigen-modes). In the second set of schemes, pre-selection at linear polarization (either p or s-states) and post-selections at linear and elliptical polarization basis are used for weak value amplification and selective observation of the IF shifts. Post selection (s) in elliptical polarization states resulted in intriguing manifestation of the two variants of the IF effect. While in one case, the two IF effect variants combine to yield partially spatial and partially angular weak value amplified IF shifts, all the shifts were converted to spatial nature leading to no observable beam shifts, for the other post selection. The demonstrated ability to amplify, controllably decouple or combine the beam shifts via weak measurements may prove to be valuable for understanding the different physical contributions of the effects and for their applications in sensing and precision metrology.